\documentclass[twocolumn,aps,prb,floatfix,nobalancelastpage]{revtex4}
\usepackage{amsmath,amssymb}
\usepackage[dvips]{graphicx}

\begin{document}

\title{Well-width dependence of valley splitting in Si/SiGe quantum wells}

\author{Kohei Sasaki,$^1$ Ryuichi Masutomi,$^1$ Kiyohiko Toyama,$^1$
Kentarou Sawano,$^2$ Yasuhiro Shiraki,$^2$ Tohru Okamoto${}^{1}$}

\affiliation{$^1$Department of Physics, University of Tokyo, 7-3-1, Hongo, Bunkyo-ku, Tokyo 113-0033, Japan\\
$^2$Research Center for Silicon Nano-Science, Advanced Research Laboratories, Tokyo City University,
8-15-1 Todoroki, Setagaya-ku, Tokyo 158-0082, Japan}

\begin{abstract}
The valley splitting in Si two-dimensional electron systems
is studied using Si/SiGe single quantum wells (QWs) with different well widths.
The energy gaps for 4 and 5.3~nm QWs,
obtained from the temperature dependence of the longitudinal resistivity 
at the Landau level filling factor $\nu=1$,
are much larger than those for 10 and 20~nm QWs.
This is consistent with the well-width dependence of the bare valley splitting
estimated from the comparison with the Zeeman splitting
in the Shubnikov-de Haas oscillations.
\end{abstract}

\maketitle
The conduction band of bulk silicon has six equivalent minima or ``valleys''
located at 85\% of the way to the Brillouin zone boundaries in the [100], [010], and [001] directions.
In two-dimensional electron systems (2DESs) 
in Si metal-oxide-semiconductor field-effect-transistors
and Si/SiGe heterostructures formed on Si(001) substrates,
the sixfold valley degeneracy is lifted to twofold due to the quantum confinement effect and the strain effect.
It has long been known that this remaining twofold degeneracy of $\pm k_z$ valleys is lifted in actual Si 2DESs,
while the mechanism is not well understood.\cite{Ando1982}
There has recently been renewed interest in the valley splitting of Si 2DESs
because of research developments in silicon-based quantum computation.\cite{Kane1998,Friesen2003,Xiao2004}
Spins in silicon are expected to have long coherence times 
due to small spin-orbit interactions and electron-nuclear spin (hyperfine) couplings.
It is believed that the valley degeneracy is a potential source of decoherence
and the valley splitting should be controlled.\cite{Koiller2002}

Low disorder silicon 2DESs are realized in high-quality strained silicon quantum wells (QWs) of 
Si/Si${}_{1-x}$Ge${}_x$ heterostructures.\cite{Schaffler1997}
Recent calculations indicate that the valley splitting in Si/SiGe heterostructures
strongly depends on the well width.\cite{Boykin2004a,Boykin2004b,Friesen2007,Valavanis2007}
However, the well-width dependence has not been studied experimentally,
while the effect of lateral confinement was investigated using a quantum point contact.\cite{Goswami2007} 
In this letter, we report magnetotransport measurements on silicon QWs with different well widths.
The results indicate that the valley splitting is much larger in narrow QWs than in wide QWs.

Four samples of well width $w=4, 5.3, 10$ and 20~nm, were grown by the molecular beam epitaxy technique.
A strained Si QW is sandwiched between relaxed Si${}_{0.8}$Ge${}_{0.2}$ layers.
The electrons are provided by a Sb-$\delta$-doped layer 20~nm above the channel.
The electron density $N_s$ can be controlled by varying back-gate voltage $V_{\rm BG}$
of a $p$-type Si(001) substrate 2.1~$\mu$m below the channel at 20~K
after brief illumination of red light emitting diode.
Sample preparation and characterization were described in detail elsewhere.\cite{Yutani1996}
Standard four-probe resistivity measurements were performed for $1.8\times 0.2$~mm${}^2$ Hall bars\cite{Toyama2008}
in a pumped ${}^3$He refrigerator. 

Figure 1(a) shows the longitudinal resistivity $\rho_{xx}$ versus 
perpendicular magnetic field $B$ of the 20~nm sample at 0.38~K.
The integer quantum Hall (QH) states are clearly observed 
for the Landau level (LL) filling factors $\nu=1$, 2 and 4.
\begin{figure}[b]
\begin{center}
\includegraphics[width=8cm]{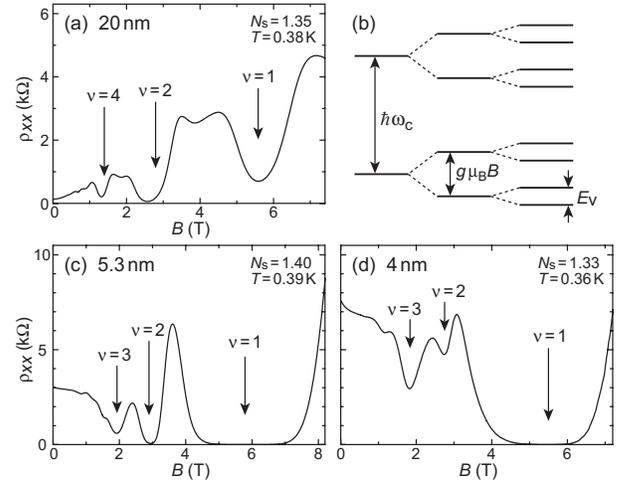}
\end{center}
\caption{
(a) Low temperature longitudinal resistivity $\rho_{xx}$ of the 20~nm sample.
The electron density is in units of $10^{15}~{\rm m}^{-2}$.
Arrows indicate magnetic fields for integer LL filling factros.
(b) Single particle LL energy diagram. LLs spaced by the cyclotron energy $\hbar \omega_c$
 are split by $g \mu_B B$ and $E_v$.
 The Fermi energy is located in the valley splitting 
 for odd integer values of $\nu=N_s h/e B$.
 For $\nu=4n$ and $\nu=4n-2$ ($n=1,2,3,...$), it is in the cyclotron gap and the Zeeman splitting, repectively.
(c) $\rho_{xx}$ of the 5.3~nm sample.
(d) $\rho_{xx}$ of the 4~nm sample.
}
\end{figure}
As shown in Fig.~1(b), 
they are related to the valley splitting $E_v$,
the Zeeman splitting $E_z=g \mu_B B$ and the cyclotron gap $\hbar \omega_c$, respectively.
Here $g$ is the $g$-factor and $\omega_c$ is the cyclotron frequency.
The $\nu=3$ QH state is not observed at this temperature.
On the other hand, the valley splitting at $\nu=3$
appears in narrower QWs as shown in Figs.~1(c) and 1(d).
In these samples, $\rho_{xx}$ reaches zero in the vicinity of $\nu=1$.

In order to study the valley splitting quantitatively,
we investigate the temperature dependence of the $\rho_{xx}$ minimum.
Typical data are shown in Fig.~2(a).
\begin{figure}[t]
\begin{center}
\includegraphics[width=7.5cm]{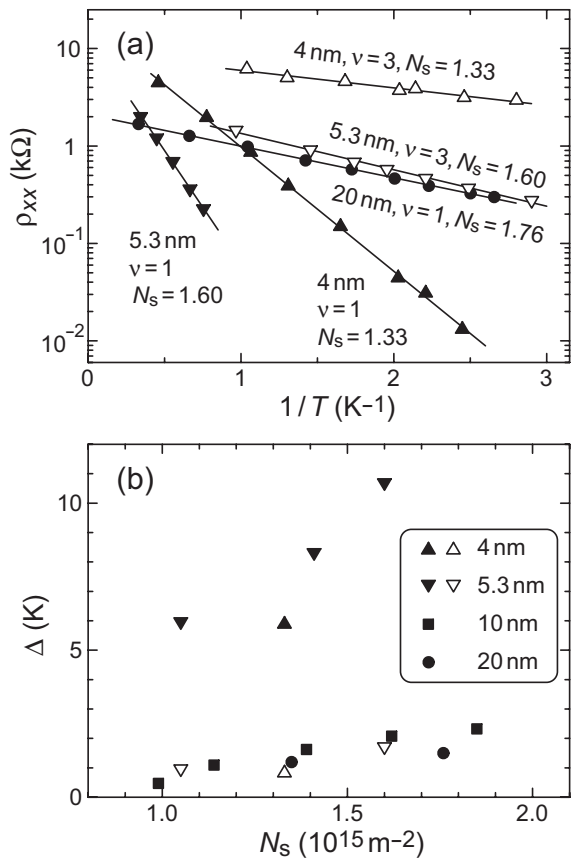}
\end{center}
\caption{
(a) Typical temperature dependence of $\rho_{xx}$ minima of the QH states
at $\nu=1$ and 3.
Arrhenius plots for different values of $w$, $\nu$ and $N_s$.
The solid lines are fits to the data.
(b) The obtained energy gap for four samples with different well widths.
The solid symbols represent the data obtained at $\nu=1$
and the open symbols are for $\nu=3$.
}
\end{figure}
The results are described well by the thermally activated form
$\rho_{xx}=\rho_0 \exp \left( -\Delta/2T \right )$,
where $\Delta$ is the the energy gap.
Note that the slope in the Arrhenius plots corresponds to $-\Delta/2$
since the chemical potential lies at the middle of the gap
for integer $\nu$ and low temperature.

In Fig.~2(b), $\Delta$ obtained for different samples is shown as a function of $N_s$.
The energy gaps of the $\nu=1$ QH state for the 4 and 5.3~nm samples 
are much larger than those for the 10 and 20~nm samples.
Recent calculations predict that
the valley splitting oscillates as a function of the well width $w$
and the oscillation amplitude decreases rapidly with $w$.\cite{Boykin2004a,Boykin2004b,Friesen2007,Valavanis2007}
The calculated values are of the order of 1~meV ($\sim 10$~K) for $w \sim 5$~nm
and of 0.1~meV ($\sim 1$~K) for $w \sim 10$~nm.
The present results for $\nu=1$ do not contradict them.
Due to modulation doping and applied gate voltage,
local electric fields are ubiquitous in heterostructures.
The average electric field in the QW region
is estimated to be about 2~MV/m in our samples.
If we use the simple approximation of Ref.~\onlinecite{Friesen2007},
the crossover of the confining potential between a square well 
and a triangular well occurs at $w \approx 9$~nm.
Thus it is expected that the effective thickness of the electron layer does not change significantly for $w>10$~nm.

As described above, the large difference in the valley gap between wide and narrow QWs is explained by the theories.
However, we should consider the effects of electron-electron Coulomb interactions
since they play crucial roles in usual 2DESs.
In the case of GaAs QH systems, the Zeeman splitting observed at $\nu=1$ or $\nu=3$ 
is an order magnitude larger than the single-particle Zeeman splitting $|g \mu_B B|$
with the bare $g$-factor of $g=-0.44$.\cite{Usher1990}
Furthermore, it is widely accepted that exchange interactions stabilize the ferromagnetic state at $\nu= {\rm odd}$
even in the limit of $g=0$.\cite{Sondhi1993,Maude1996}
For small $g$, the lowest-energy charged excitations at $\nu=1$ are spin-texture excitations
knows as Skyrmions.\cite{Sondhi1993}
In the Hartree-Fock (HF) approximation for ideal (zero-thickness) 2DESs with wave functions projected onto a single LL, 
the gap for $g=0$ is given by
\begin{eqnarray}
\Delta_{\rm SK} = \frac{1}{2}\sqrt{\frac{\pi}{2}}\frac{e^2}{4\pi\epsilon_0\kappa_{\rm sc} \ell_B},
\end{eqnarray}
where $\kappa_{\rm sc}$ is the dielectric constant in the semiconductor
and $\ell_B \equiv (\hbar /eB)^{1/2}=(\nu/2\pi N_s)^{1/2}$ is the magnetic length.
This is exactly half the exchange energy cost of a single spin-flip excitation.
When Eq.~(1) is applied to the valley (pseudospin) splitting in Si 2DESs,
it is calculated to be $\Delta_{\rm SK} = 88$~K for $\nu=1$
and $\Delta_{\rm SK} = 51$~K for $\nu=3$
at $N_s=1.5 \times 10^{15}~{\rm m}^{-2}$.
Note that we consider the lowest Landau orbital even for $\nu=3$ in Si 2DESs.
While Eq.~(1) gives only the lowest exchange energy cost,
these values are much larger than the experimental ones.
The discrepancy between Eq.~(1) and experiment is well known
for the Zeeman splitting in GaAs QH systems,
and discussed in relation to the effects of disorder, finite thickness,
and LL mixing.\cite{Maude1996,Schmeller1995,Shukla2000}
Disorder broadening of LLs is expected to be strong
in low mobility samples.
In our case, the zero-field mobility $\mu$ decreases rapidly with decreasing $w$.
The 4~nm sample has $\mu=0.62~{\rm m}^{2}/{\rm Vs}$ at $N_s=1.33 \times 10^{15}~{\rm m}^{-2}$.
The half width of the level broadening $\Gamma=1/2\tau$ is calculated to be 5.7~K,
where $\tau$ is the scattering time.
As shown in Fig.~2(b), the measured gap for $w=4$~nm is slightly smaller than that for $w=5.3$~nm.
This may be caused by the disorder broadening effect.
However, it does not account for the observed small gaps of the 10~nm and 20~nm samples
with much higher mobility.
The finite thickness of the electron layer leads to the softening of the Coulomb interaction
and reduces $\Delta_{\rm SK}$.\cite{Cooper1997,Fertig1997}
However, the correction for Eq.~(1) is calculated to be small in our QWs.\cite{HFvalues}
LL mixing may be the most important effect for our samples.
In Ref.~\onlinecite{Dickmann2002},
it is shown that the Skyrmion gap is substantially reduced in comparison with the HF calculations
when the ratio of the typical Coulomb energy 
$E_C={e^2}/{4\pi\epsilon_0\kappa_{\rm sc} \ell_B}$ to $\hbar \omega_c$
becomes comparable with or larger than unity.
In Si 2DESs, the ratio is given by
$r_C=3.9 (\nu/N_s[10^{15}~{\rm m}^{-2}])^{1/2}$.
Thus the effect of LL mixing is expected to be strong in our samples.
As $B$ decreases, the Coulomb energy decreases and the effect of LL mixing increases.
This explains the observed $N_s$ dependence of $\Delta$ and
the large difference between $\nu=1$ and $\nu =3$.

To obtain the bare valley splitting $E_v$ from the measured gap $\Delta$,
we need to evaluate the exchange enhancement.
To our knowledge, however, there is no available theory for large $r_C$ at present.
Instead, here we estimate the bare valley splitting from the comparison with the Zeeman splitting
in the Shubnikov-de Haas oscillations.
We assume that $E_v$ does not depend on $B$ and $N_s$,
and the exchange enhancement and the disorder smearing occur
equally for the valley and Zeeman splittings.
For the 4~nm sample, the $\nu=3$ QH state
is stronger than the $\nu=2$ QH state, as shown in Fig.~1(d),
although the enhancement is expected to be smaller for larger $\nu$.
Thus the bare valley splitting $E_v$ should be larger than $g\mu_B B-E_v$.
This leads to $E_v > g\mu_B B/2=1.8$~K with $g=+2.0$ and $B=2.7$~T.
On the other hand,
we do not expect a crossing between the second and third LLs at $\nu=2$
since the QH state becomes stronger as the magnetic field is tilted away from normal to the 2D plane (not shown here).
This indicates $E_v < g\mu_B B=3.6$~K.
For the 5.3~nm sample, we were able to compare the $\nu=2$ and $\nu=3$ QH states
at the same $B$ by adjusting $N_s$.
From the measured gaps, we obtain $E_v=1.4\pm 0.4$~K taking into account small $N_s$-dependence of the mobility.
For the 20-nm sample, some of the present authors have studied the Shubnikov-de Haas oscillations
in low magnetic fields below 0.1~K.\cite{Toyama2008}
The Zeeman splitting can be observed even at $B=0.29$~T ($\nu=34$)
while the valley splitting is smeared out.
Thus the bare valley splitting in this sample is estimated to be less than 0.2~K.
Using similar procedures, we obtain $E_v < 0.7$~K for the 10~nm sample.
The results are summarized in Table I.
Although there is some uncertainty,
it is confirmed that the bare valley splitting increases
rapidly with decreasing well width.

\begin{table}
\begin{ruledtabular}
\caption[TABLE I]{Bare valley splitting estimated from the Shubnikov-de Haas oscillations.
The average electric field $\bar{F}$ in the QW region at the corresponding back-gate voltage
is also presented for future calculations.} 
\label{table2}
\begin{tabular}{ccc}
 $w$ & $E_v$ & $\bar{F}$\\
 $({\rm nm})$ & $({\rm K})$ & $({\rm MV/m})$\\
\colrule
 4 & $1.8 - 3.6$ & 1.0 \\
 5.3 & $1.4 \pm 0.4$ & 1.3 \\
 10 & $<0.7$ & 2.0 \\
 20 & $<0.2$ & 2.3 \\
\end{tabular}
\end{ruledtabular}
\end{table}

In summary, 
we have studied the valley splitting in Si/SiGe heterostructure samples with different well widths.
The energy gap obtained from the $T$-dependence of $\rho_{xx}$
is much larger in narrow QWs than in wide QWs,
while the degree of the exchange enhancement is unknown at present.
The bare valley splitting estimated from the comparison with the Zeeman splitting
also exhibits a rapid increase with decreasing well width.

This work has partly supported by Grant-in-Aids for Scientific Research (A) (Grant No. 21244047)
and Grant-in-Aid for Scientific Research on Priority Area
``Physics of New Quantum Phases in Superclean Materials'' (Grant No. 20029005) from MEXT, Japan.


\begin{thebibliography}{1}
\bibitem{HFvalues}
Using the simple approximation for the confining potential of Ref.~\protect\onlinecite{Friesen2007},
$\Delta_{\rm SK}$ for $\nu=1$ at $N_s=1.5 \times 10^{15}~{\rm m}^{-2}$ is obtained to be
78~K, 75~K, 68~K and 67~K, 
for QWs with $w=4, 5.3, 10$ and 20~nm, respectively, from HF calculations (Ref.~\protect\onlinecite{Cooper1997}).
\end{thebibliography}
\end{document}